\documentclass[conference]{IEEEtran}
\IEEEoverridecommandlockouts
\usepackage{makecell}
\usepackage{diagbox}
\usepackage{amsmath,amssymb,amsfonts}
\usepackage{algorithmic}
\usepackage{graphicx}
\usepackage{textcomp}
\usepackage{xspace}
\usepackage{booktabs}
\usepackage{xcolor}
\usepackage{cite}
\usepackage[bookmarks=true,breaklinks=true,letterpaper=true,colorlinks,linkcolor=blue,citecolor=magenta,urlcolor=blue]{hyperref}

\def\BibTeX{{\rm B\kern-.05em{\sc i\kern-.025em b}\kern-.08em
    T\kern-.1667em\lower.7ex\hbox{E}\kern-.125emX}}

\ifx\figurename\undefined \def\figurename{Figure}\fi
\renewcommand{\figurename}{Fig.}
\renewcommand{\paragraph}[1]{\textbf{#1} }

\newcommand{\Sect}[1]{Sec.~\ref{#1}}
\newcommand{\Fig}[1]{Fig.~\ref{#1}}
\newcommand{\Tbl}[1]{Tbl.~\ref{#1}}

\newcommand{\proj}{\textsc{StreamingGS}\xspace}

\begin{document}



\title{\proj: Voxel-Based Streaming 3D Gaussian Splatting with Memory Optimization and Architectural Support}

\author{
    Chenqi Zhang\IEEEauthorrefmark{1*}, 
    Yu Feng\IEEEauthorrefmark{1*}, 
    Jieru Zhao\IEEEauthorrefmark{1$\dagger$}, 
    Guangda Liu\IEEEauthorrefmark{1}, 
    Wenchao Ding\IEEEauthorrefmark{2}, 
    Chentao Wu\IEEEauthorrefmark{1},
    Minyi Guo\IEEEauthorrefmark{1}  \\
    \IEEEauthorblockA{
        \IEEEauthorrefmark{1} School of Computer Science, Shanghai Jiao Tong University,  
        \IEEEauthorrefmark{2} Academy for Engineering \& Technology, Fudan University  
    }
    \IEEEauthorblockA{
       \IEEEauthorrefmark{} \{zhangchenqi123, y-feng, zhao-jieru\}@sjtu.edu.cn 
       \IEEEauthorrefmark{*}Equal contribution. \IEEEauthorrefmark{$\dagger$}Corresponding author. \
    }
    
    \thanks{This paper has been accepted by the Design Automation Conference (DAC) 2025.}
}

\maketitle

\begin{abstract}

3D Gaussian Splatting (3DGS) has gained popularity for its efficiency and sparse Gaussian-based representation. 
However, 3DGS struggles to meet the real-time requirement of 90 frames per second (FPS) on resource-constrained mobile devices, achieving only  2 to 9 FPS.
Existing accelerators focus on compute efficiency but overlook memory efficiency, leading to redundant DRAM traffic. 
We introduce \proj, a fully streaming 3DGS algorithm-architecture co-design that achieves fine-grained pipelining and reduces DRAM traffic by transforming from a tile-centric rendering to a memory-centric rendering.
Results show that our design achieves up to 45.7$\times$ speedup and 62.9$\times$ energy savings over mobile Ampere GPUs.

\end{abstract}


\section{Introduction}
\label{sec:intro}



In virtual and augmented reality (VR/AR), photo-realistic and real-time rendering is essential for delivering immersive user experiences. 
Among various rendering techniques~\cite{pharr2023physically, deng2017toward, mildenhall2021nerf, barron2021mip, wu2024recent, chen2024survey}, 3D Gaussian Splatting (3DGS)~\cite{kerbl20233d, fang2024mini, lee2023compact, fan2023lightgaussian, lin2025metasapiens, huang2025seele} has quickly gained its popularity for its efficient rendering process and sparse Gaussian representation (also called Gaussian ellipsoids) of scene geometries.

Nevertheless, 3DGS still falls short of the real-time requirement in the VR/AR domain, namely 90 frames per second (FPS)~\cite{questprospec, visionprospec, wang2023effect}, on mobile devices like VR headsets. 
Our experiments show that 3DGS merely achieves 2 to 9 FPS on the Nvidia Orin NX~\cite{orinsoc}~\footnote{The performance of the Ampere GPU (3.7 TFLOPS, released in February 2023) is comparable to that of the GPU (3.5 TFLOPS) in the latest Snapdragon XR2~\cite{snapdragonxr2}, released in September 2023 for VR.}. 
Although several studies have proposed dedicated accelerators for 3DGS~\cite{lee2024gscore}, their designs primarily optimize compute efficiency and often overlook off-chip memory traffic during rendering, making them hard to satisfy the real-time requirement in real-world scenarios. 

\begin{figure}
    \centering
    \includegraphics[width=\linewidth]{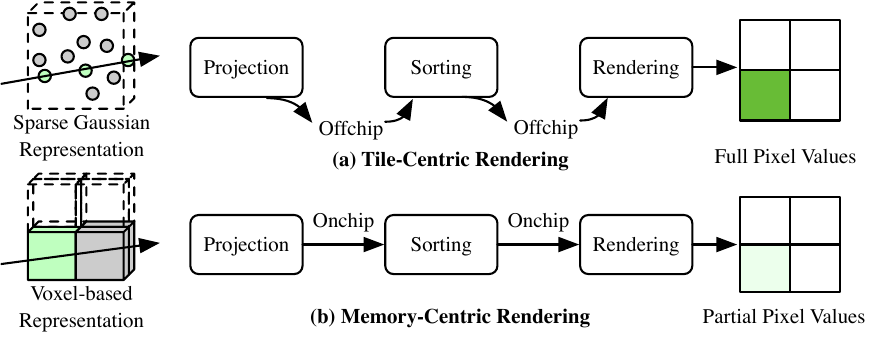}
    \caption{Tile-centric rendering (a) vs. memory-centric rendering (b). 
    The computation pattern is highlighted in green.
    Tile-centric rendering computes the complete pixel values tile-by-tile.
    Memory-centric rendering computes the partial pixel values voxel-by-voxel and regularizes the memory accesses.
    }
    \label{fig:intro}
\end{figure}

Algorithmically, existing methods~\cite{kerbl20233d, lee2024gscore} focus on a tile-centric rendering paradigm which requires excessive intermediate data communication between stages, leading to frequent off-chip traffic in \Fig{fig:intro}a.
Our characterization shows that, under a 90 FPS real-time requirement, the DRAM traffic of the current rendering paradigm would exceed the bandwidth limits of today's mobile devices on real-world scenes (\Sect{sec:bm:mot}).

The high data communication in the conventional tile-centric rendering comes from sequential processing ray-Gaussian intersections for each individual pixel, leading to redundant off-chip traffic and memory irregularity~\cite{feng2024potamoi, lee2024gscore, li2024celeritas, feng2024cicero}.
To address the bandwidth limitation, we argue to shift the rendering paradigm from a \textit{tile-centric} to a \textit{memory-centric}  approach. 
Here, we introduce a fully streaming algorithm for 3DGS to address memory access irregularity (\Sect{sec:algo}).
Our key idea is to regularize sparse Gaussian representations by partitioning the scene into voxels and performing on a voxel basis as shown in \Fig{fig:intro}b.
With all the points within a voxel stored contiguously in DRAM, we allow all points within a voxel to be streamed onto the on-chip buffer and only write the final pixel values off-chip, completely eliminating the intermediate off-chip traffic between stages.



While our memory-centric approach eliminates the intermediate off-chip traffic, our approach potentially introduces unnecessary off-chip traffic at the voxel streaming, as all points within a voxel are loaded on-chip as a whole.
To reduce the unnecessary off-chip traffic, we propose hierarchical filtering to progressively exclude irrelevant Gaussians (\Sect{sec:algo:fs}). 

Specifically, we split each Gaussian feature into two halves and intentionally keep the first half lightweight.
We then perform coarse yet lightweight computation to filter out irrelevant points based on the first half (i.e., the position and radius). 
Next, we perform precise filtering to further remove the irrelevant Gaussians by loading the second half. 
To further reduce the overhead of feature loading, we compress the second half using vector quantization.
By storing a compact codebook on-chip, we just need to fetch their lightweight indices from DRAM. 
In this way, we drastically reduce the DRAM traffic at the voxel streaming.



To unleash the full potential of our fully streaming algorithm, we co-design an efficient accelerator (\Sect{sec:arch}), including 
a voxel sorting unit to accelerate the determination of voxel rendering order and a dedicated hierarchical filtering unit to support Gaussian filtering at the voxel streaming.
Coupled with our algorithmic optimizations, our design achieves a 45.7$\times$ speedup and 62.9$\times$ energy savings compared to a Nvidia mobile Ampere GPU~\cite{orinsoc}.
Even compared against the state-of-the-art accelerator, GSCore~\cite{lee2024gscore}, we can achieve a 2.1$\times$ speedup and 2.3$\times$ energy savings.


In summary, our contributions are as follows:
\begin{itemize}
    \item We introduce a \textit{memory-centric} rendering paradigm that can achieve fully streaming 3DGS rendering, completely avoiding intermediate off-chip traffic. 
    \item We propose a hierarchical filtering scheme to further reduce the data traffic of the voxel streaming.
    \item With our architectural supports, our optimizations achieve an order of magnitude speedup and energy savings compared to a mobile Ampere GPU. 
    We also achieve a 2.1$\times$ speedup and 2.3$\times$ energy savings against GSCore.
\end{itemize}

\vspace{-0.2cm}

\section{Background and Characterization}
\vspace{-0.2cm}

\label{sec:bm}

\begin{figure}
    \centering
    \includegraphics[width=\linewidth]{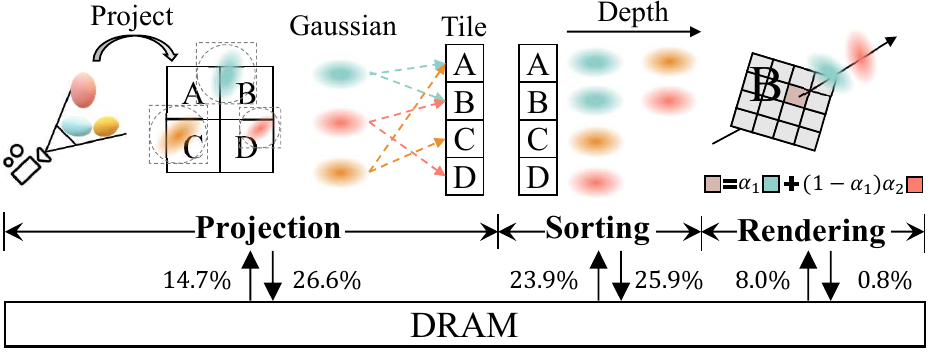}
    \caption{The overall rendering pipeline of 3DGS. 
    The percentage numbers represent the DRAM traffic proportion across stages. 
    }
    \label{fig:origin_gs}
\end{figure}


\subsection{Background}
\label{sec:bm:bg}

Overall, 3DGS uses Gaussian ellipsoids to render a scene in a tile-by-tile fashion. 
\Fig{fig:origin_gs} shows that this process can be classified into three stages:
\begin{itemize}
    \item \paragraph{Projection.} 
    This stage projects each Gaussian ellipsoid onto the 2D screen and determines its intersected tiles as shown on the left side of \Fig{fig:origin_gs}.
    Each Gaussian might intersect with multiple tiles, as the colored dashed arrow lines show.
    Meanwhile, this stage calculates Gaussian properties, such as color and depth, for the later stages. 
    \item \paragraph{Sorting.} Each tile then sorts its intersected Gaussians based on Gaussian depths to ensure the correct rendering order from the closest to the farthest.
    \item \paragraph{Rendering.} Every pixel within a tile traverses the same sorted Gaussian list and accumulates Gaussian colors to itself via alpha blending. 
    For instance, we show tile $B$ needs to blend two Gaussians, a blue one and a red one. 
\end{itemize}

\subsection{Characterization}
\label{sec:bm:mot}

\begin{figure}[t]
\begin{minipage}[t]{0.48\linewidth}
        \centering
        \includegraphics[width=\textwidth]{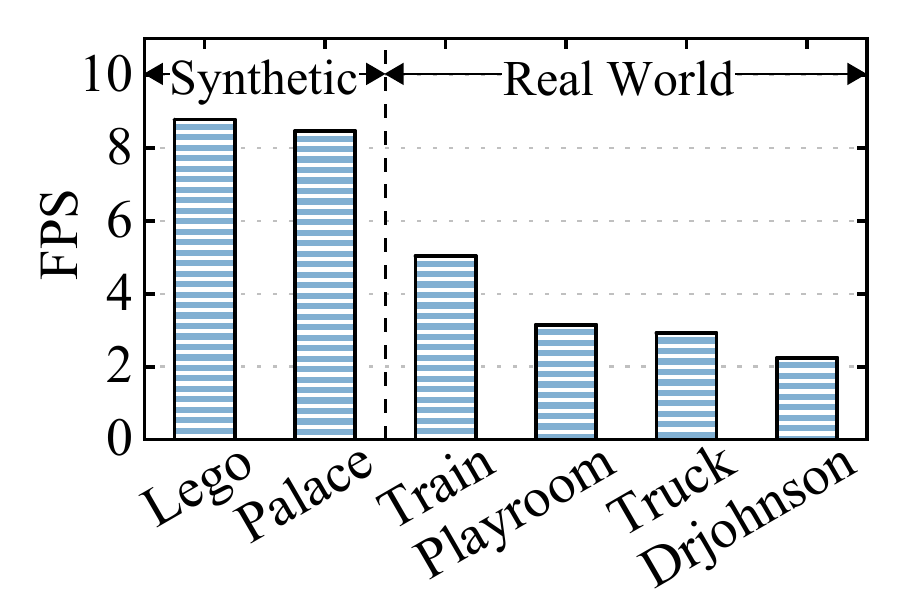}
        \caption{FPS of different scenes on a mobile SoC, Orin NX. 3DGS cannot achieve real-time on both synthetic and real-world scenes.}
        \label{fig:fps}
    \end{minipage}
    \hspace{2pt}
    \begin{minipage}[t]{0.48\linewidth}
        \centering
        \includegraphics[width=\textwidth]{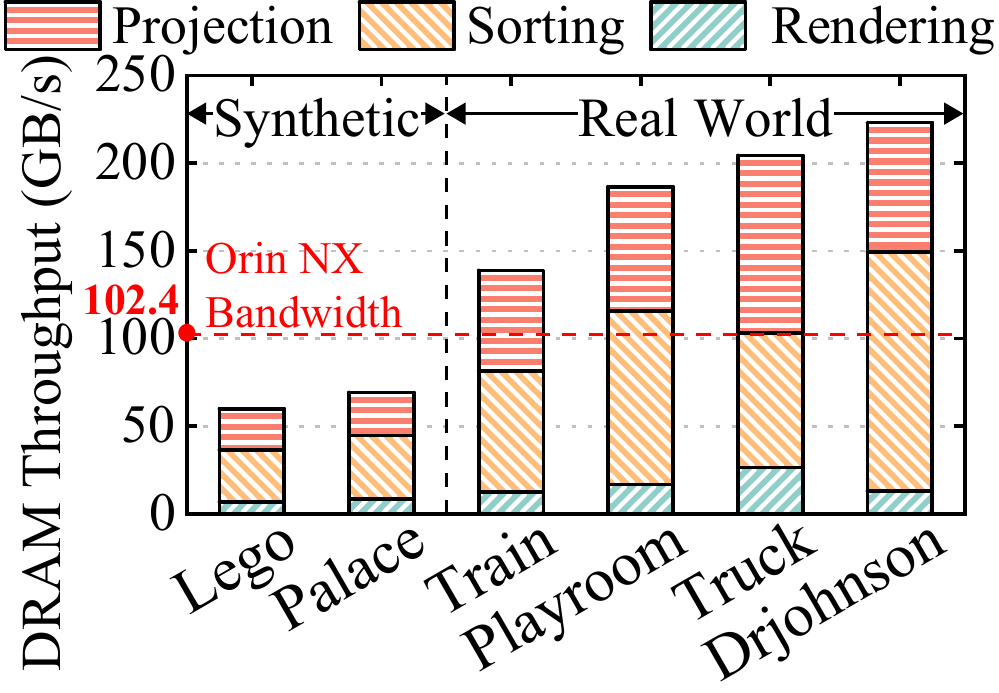}
        \caption{The DRAM bandwidth requirement for 90 FPS under different scenes. The red dashed line highlights the bandwidth limit of Orin NX.}
        \label{fig:bw}
    \end{minipage}
\end{figure}

\paragraph{Performance.} 
We first show that 3DGS algorithms cannot achieve real-time performance on mobile devices. 
\Fig{fig:fps} characterizes the FPS of 3DGS~\cite{kerbl20233d} across various scenes on NVIDIA’s Jetson Orin NX~\cite{orinsoc}. 
As the rendering scenes become more complex, shifting from synthetic to real-world scenes, performance drops from 8.5 to 4.9 FPS.

\paragraph{DRAM Traffic.}
We then show that DRAM traffic has a strong correlation with performance. 
\Fig{fig:bw} illustrates the DRAM throughput required to achieve the real-time requirement (90 FPS) for AR/VR applications~\cite{questprospec, visionprospec} across different rendering scenes. 
For real-world scenes, the DRAM bandwidth demand exceeds the bandwidth limit of a recently released mobile device, Nvidia Orin NX~\cite{orinsoc}, as highlighted in the red dashed line.
This means that the data communication requirement alone has made it impossible to achieve real-time.

We further dissect the DRAM traffic contributions across different scenes.
As shown in \Fig{fig:bw}, \textit{projection} and \textit{sorting} dominate the overall off-chip data communication, together accounting for 90\% of the total DRAM traffic.
The reasons that \textit{projection} and \textit{sorting} become the major contributors are:
\begin{itemize}
    \item In projection,  each Gaussian requires loading 59 parameters into the on-chip buffer and writing the processed features and intersection metadata back to DRAM, accounting for 41\% of DRAM traffic.
    \item In sorting, repetitive read and write operations for progressive Gaussian sorting lead to repetitive DRAM access, taking up over 49\% of DRAM traffic.
\end{itemize}

Overall, the intermediate data communication accounts for 85\% of the total off-chip traffic in \Fig{fig:origin_gs}.
This result motivates us to propose a fully-streaming algorithm that minimizes the off-chip traffic.
\Sect{sec:algo} presents our algorithm design and \Sect{sec:arch} explains the architectural support for our algorithm.

\section{Algorithm Design}
\label{sec:algo}

\Sect{sec:algo:overview} first gives an algorithm overview and challenges that we need to address. 
Then, \Sect{sec:algo:fs} and \Sect{sec:algo:dc} explain the key components in our algorithm and the data layout that supports our algorithm, respectively.

\begin{figure*}
    \centering
    \includegraphics[width=\linewidth]{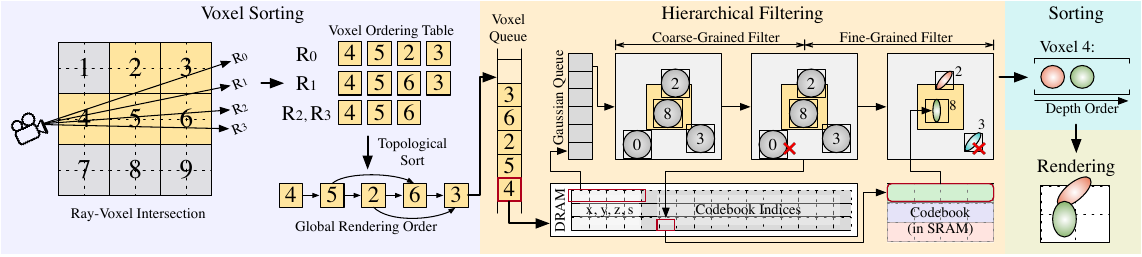}
    \caption{The overview of our fully-streaming algorithm. Our algorithm initially partitions the entire 3D scene into small voxels. For each image tile, we first establish a rendering order of the intersected voxels via topological sort, then, process these voxels sequentially. During each voxel processing, we apply hierarchical filtering to further reduce the Gaussian loading within a voxel before the remaining sorting and rendering stages. The final pixel values are obtained by accumulating all the partial results across voxels.}
    \label{fig:overview}
\end{figure*}



\begin{figure}[t]
\begin{minipage}[t]{0.49\linewidth}
        \centering
        \includegraphics[width=\textwidth]{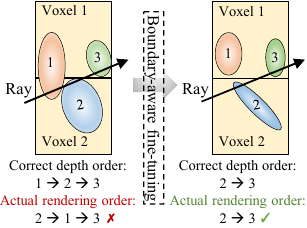}
        \caption{An example of incorrect rendering order among Gaussians and the effect of boundary-aware fine-tuning.}
        \label{fig:incorrect_depth_example}
    \end{minipage}
    \hspace{2pt}
    \begin{minipage}[t]{0.48\linewidth}
        \centering
        \includegraphics[width=\textwidth]{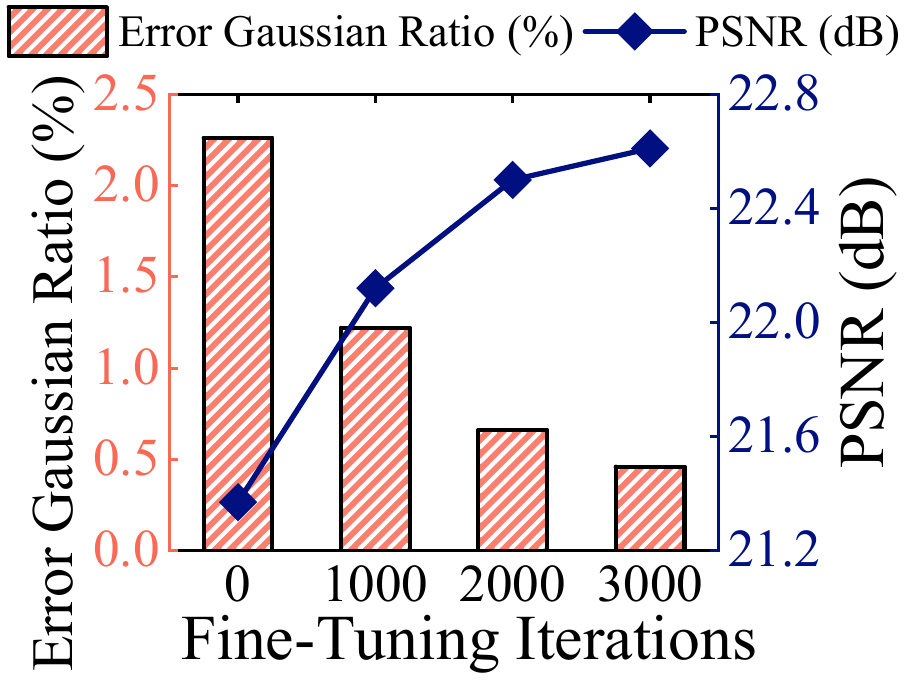}
        \caption{Changes in the ratio of Gaussians with incorrect depth order and PSNR during the boundary-aware fine-tuning process.}
        \label{fig:incorrect_depth_FT}
    \end{minipage}
\end{figure}

\subsection{Overview}
\label{sec:algo:overview}

The main downside of the original 3DGS pipeline is that it renders a frame tile-by-tile and each tile executes three stages sequentially.
The reason is that the intermediate data of each stage is too large to be accommodated on-chip, inevitably, introducing frequent off-chip traffic during the execution.

\paragraph{Idea.} To eliminate frequent off-chip communications at intermediate stages, our idea is to partition an entire 3D scene into small voxels and incrementally render the entire scene voxel-by-voxel.
By processing one voxel at a time, our algorithm guarantees the intermediate data between stages can be completely held on-chip, thus, no off-chip traffic occurs.

\Fig{fig:overview} gives an overview of our algorithm.
Our algorithm renders a group of pixels together, e.g., $R_0$ to $R_3$ are in the same group.
For each group, we first identify which voxels intersect with these pixel rays and store the results in the voxel rendering table.
Once each pixel's intersected voxels are determined, we establish their global rendering order. 
Then, each voxel goes through hierarchical filtering, sorting, and rendering in order, and obtain the partial pixel values of this tile.
The final pixel values are obtained by accumulating all the partial results across voxels.
By rendering one voxel at a time, each stage requires minimal data communication, thus, no off-chip traffic is required throughout the execution.

\paragraph{Challenges.} 
However, there are three challenges when applying our algorithm:
\begin{itemize}
    \item \underline{\textit{Inter-Voxel Order.}} Pixels within a single group often intersect with multiple voxels, as shown in \Fig{fig:overview}. 
    The rendering order for each pixel should be from the closest voxel to the farthest. 
    The first challenge is to ensure the correct global order for all intersected voxels.
    \item \underline{\textit{Cross-Voxel Gaussian Order.}}
    Recall, each pixel is required to render Gaussians from the closest to the farthest in 3DGS, but the correct voxel-level ordering does not always guarantee the rendering order correctness at Gaussian-level granularity.
    For instance, on the left side of \Fig{fig:incorrect_depth_example}, Gaussian 2 in voxel 2 renders before Gaussian 1 in voxel 1, whereas the correct order should be reversed.
    \item \underline{\textit{Redundant Gaussians in Voxels.}}
    Lastly, each voxel includes Gaussians within close proximity; however, loading an entire voxel might unintentionally bring unrelated Gaussians on-chip. Leaving these redundant Gaussian without care would introduce excessive DRAM traffic.
\end{itemize}





\vspace{-3pt}

\subsection{Fully Streaming Algorithm}
\label{sec:algo:fs}

In this section, we explain how we address these challenges.

\paragraph{Inter-Voxel Order.} 
To establish the correct voxel order across pixels in each tile, we first collect the rendering order of individual pixels in a voxel ordering table in \Fig{fig:overview}. 
Based on individual rendering orders, we construct a direct acyclic graph (DAG), where each voxel is represented as a node, and edges represent the rendering dependencies.
We then apply a topological sort~\cite{kahn1962topological} on the DAG to guarantee that no cycles in this graph and all the render dependencies are satisfied.
The sorted order is then used in the remaining process.

\paragraph{Cross-Voxel Gaussian Order.}
To ensure correct rendering order among Gaussians, we observe that the incorrect order occurs only when a Gaussian spans across multiple voxels. 
In \Fig{fig:incorrect_depth_example}, Gaussians should be rendered from Gaussian 1 (the closest) to Gaussian 3 (the farthest). 
However, with voxel-by-voxel rendering, Gaussian 2 is incorrectly rendered before Gaussian 1.
\Fig{fig:incorrect_depth_FT} shows that incorrect ordering can lead to a degradation in rendering quality.

To alleviate this issue, we introduce a \textit{boundary-aware fine-tuning} to adjust cross-voxel Gaussians.
Specifically, we keep each Gaussian position fixed to retain the scene geometry while fine-tuning its remaining trainable parameters, such as scale, orientation, etc. 
The fine-tuning reduces the size and modifies the orientation of Gaussians, minimizing their likelihood of spanning across other voxels. 



In our fine-tuning, we propose a new loss function to punish the cross-boundary Gaussians,
\begin{equation}
\mathcal{L}=\mathcal{L}_{origin} + \beta \mathcal{L}_{CBP},
\end{equation}
\noindent where $\mathcal{L}_{origin}$ is the original loss in 3DGS and $\mathcal{L}_{CBP}$ is the cross-boundary penalty loss which penalizes Gaussians that span across voxel boundaries.
$\beta$ is a hyper-parameter to balance the influence of $\mathcal{L}_{CBP}$. 

$\mathcal{L}_{CBP}$ minimizes the scale of the Gaussians that cross voxel boundaries and is expressed as,
\begin{align}
    \mathcal{L}_{CBP} & = \frac{1}{N} \sum_{i=0}^{N-1}S_iT_i, \\
    \text{where}\ T_i & =
        \begin{cases} 
            1, & \text{if } \text{depth}_i < \max_{j \in [0, i)} \{\text{depth}_j\}, \\ \nonumber
            0, & \text{otherwise},
        \end{cases}                             
\end{align}
\noindent where N is the total number of Gaussians, $S_i$ is the maximum scale of Gaussian i. 
$T_i$ is an indicator that checks if current Gaussian $i$ has the largest depth compared to all previously rendered Gaussians.
If this condition is not satisfied, we penalize it by setting $T_i$ to be $1$.



\Fig{fig:incorrect_depth_FT} illustrates the effectness of our boundary-aware fine-tuning. 
After 3,000 iterations, the percentage of cross-boundary Gaussians decreases from 2.3\% to 0.4\%, while the rendering quality regains from 21.37~dB to 22.61~dB.

\paragraph{Redundant Gaussians in Voxels.}
Recall that during projection, each Gaussian requires 59 parameters.
To avoid irrelevant Gaussians, we propose a two-phase \textit{hierarchical filtering} to progressively fetch Gaussian parameters in \Fig{fig:overview}.

The first phase serves as a lightweight but coarse-grained filter that approximates whether Gaussians are intersected with the image tile. 
In this phase, each Gaussian only requests 4 parameters (3D coordinates and maximum scale) from off-chip memory to compute the projected center and the maximum projected radius with respect to the image tile.
If a Gaussian does not pass the coarse-grained filter, we skip its remaining computations and avoid loading its remaining 55 parameters.
For example, Gaussian $0$ in \Fig{fig:overview} does not intersect with the image tile in yellow, we can safely remove it from the subsequent computation. 

In the second fine-grained filtering, we identify the actual Gaussians that are intersected with the image tile by computing precise projection.
For example, after getting shapes and orientations, we can know that Gaussian $3$ in \Fig{fig:overview} does not intersect with the image tile actually and hence can skip its subsequent sorting and rendering. 
To further reduce the off-chip traffic, we propose a novel data compression scheme to encode the remaining parameters into a compact codebook.
This way, we only need to load codebook indices from DRAM instead of the original parameters.
We explain our data compression scheme in \Sect{sec:algo:dc}.
Combining coarse- and fine-grained filtering, our hierarchical filtering reduces the number of Gaussians processed in each voxel by 76.3\%.


\subsection{Customized Data Layout}
\label{sec:algo:dc}

\begin{figure}
    \centering
    \includegraphics[width=\linewidth]{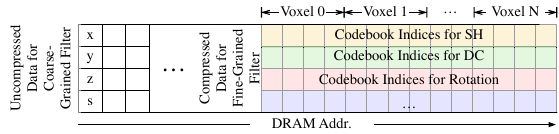}
    \caption{The DRAM Data layout. We separate the Gaussian features into two halves for two filtering phases and compress the data for the second phase using vector quantization to reduce the off-chip communication.}
    \label{fig:layout}
\end{figure}

To reduce data communication in hierarchical filtering, we propose a customized data layout in \Fig{fig:layout}.
Overall, we divide all Gaussian features into two halves, which are stored separately in DRAM and used at different filtering phases.
The first half, containing 4 parameters (the 3D coordinates and maximum scale), is used for the coarse-grained filter. 
The second half, used in the fine-grained filter, includes the remaining parameters, such as rotation matrices.

To alleviate DRAM traffic, we apply the vector quantization to compress the second half, so that we store only codebook indices off-chip while keeping the codebook itself on-chip for decoding. 
In addition, we encode different parameters into separate codebooks to preserve quantization precision and adopt the quantization-aware fine-tuning~\cite{lee2023compact} to update the Gaussian parameters, allowing the indices to capture feature variations without loss of detail.
Our result shows that, during the voxel streaming, only the codebook indices are fetched from DRAM and reduce 92.3\% DRAM traffic.
Note that, we only compress the second half, because we empirically find that compressing the first half leads to accuracy loss.

\vspace{-3pt}
\section{Architectural Support}
\label{sec:arch}

\begin{figure}
    \centering
    \includegraphics[width=\linewidth]{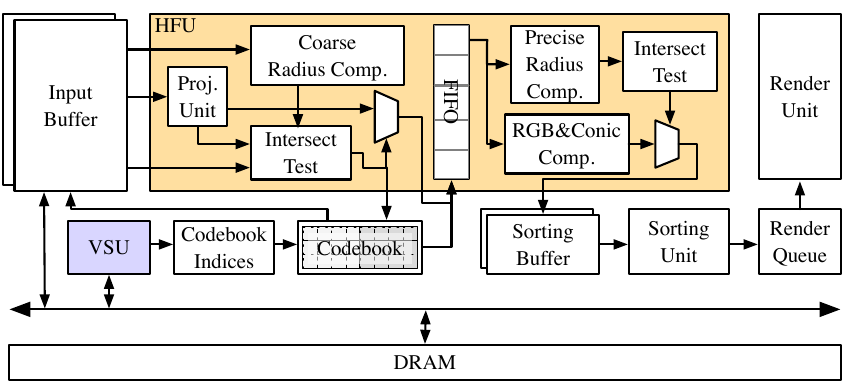}
    \caption{Overview of \proj accelerator, consisting of four components: a voxel sorting unit (VSU), a hierarchical filtering unit (HFU), a sorting unit, and a rendering unit. Our contributions are highlighted in colors: HFU is in yellow and VSU is in purple. The detailed design of VSU is in \Fig{fig:vsu}.}
    \label{fig:arch}
\end{figure}

\subsection{Overview} 
\label{sec:arch:ov}

To support our fully-streaming algorithm, we propose our co-designed accelerator in \Fig{fig:arch}.
Our architecture consists of four components: a voxel sorting unit (VSU), a hierarchical filtering unit (HFU), a sorting unit, and a rendering unit.

Given a set of ray directions, VSU first identifies and sorts the intersected voxels.
Then, for each intersected voxel, HFU applies a hierarchical filter to identify intersected Gaussians.
Both designs will be explained in \Sect{sec:arch:vsu} and \Sect{sec:arch:hfu}, respectively.
Once HFU obtains the filtered Gaussians, sorting and rendering units execute the sorting and rendering stage from their individual buffer, respectively.
Both sorting and rendering designs are adopted from GSCore~\cite{lee2024gscore}.
However, we simplify the sorting unit by just adopting the bitonic sorting unit from GSCore, as our voxel-based rendering only requires establishing the rendering order for Gaussians within a voxel.


\begin{figure}
    \centering
    \includegraphics[width=\linewidth]{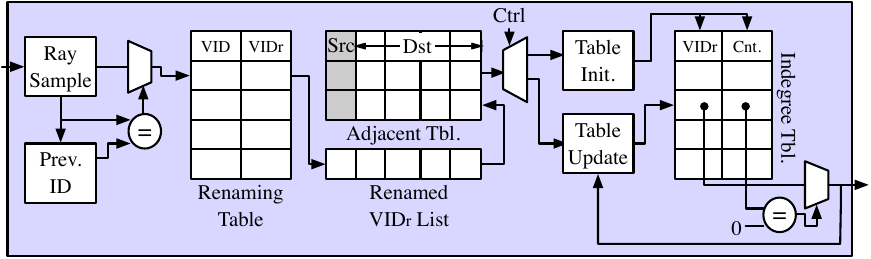}
    \caption{The design of voxel sorting unit (VSU). It first identifies the intersected voxels for every pixel within a tile and then establishes the global rendering order of these voxels for sorting and rendering.}
    \label{fig:vsu}
\end{figure}

\subsection{Voxel Sorting Unit (VSU)}
\label{sec:arch:vsu}

The architecture of VSU is illustrated in Fig. \ref{fig:vsu}.  
For pixels within a group, VSU samples along each pixel ray to identify the intersected voxels and performs a topological sort to establish the global rendering order of these voxels.
Since we partition the entire scene into voxels offline, the voxel ID (VID) can be directly computed by the coordinate of a ray sample.
To reduce the VID range, we further filter out those empty voxels with no Gaussians by augmenting a renaming table in VSU to reassign each VID with a new VID$_\text{r}$.

Each time we obtain a renamed VID$_\text{r}$ list for one pixel, we use this list to construct an adjacent table.
This adjacent table functions like a cache to help the dependency lookup among voxels.
Each table entry has a tag and a value, representing a source voxel and all its corresponding destination voxels, respectively.
The adjacent table is constructed once VSU finishes iterating all pixel rays in a group.

VSU then uses the adjacent table to initialize an in-degree table.
Each entry in the in-degree table is directly indexed by a VID$_\text{r}$ value and stores its in-degree count, which records the number of source voxels this VID$_\text{r}$ has.
If an in-degree of a VID$_\text{r}$ becomes zero, this VID$_\text{r}$ has no source dependency and can be written to the later stage.
Each time VSU outputs a VID$_\text{r}$, VSU also updates the in-degree table accordingly.
The topological sort finishes when the in-degree table is empty.

\vspace{-0.1cm}

\subsection{Hierarchical Filtering Unit (HFU)}
\label{sec:arch:hfu}

Once VSU generates the rendering order of intersected voxels, HFU starts to load one voxel at a time and process the Gaussians within this voxel. 
Overall, HFU, highlighted in yellow in \Fig{fig:arch}, consists of a frontend and a backend, responsible for coarse-grained and fine-grained filtering, respectively.


\paragraph{Coarse-Grained Filter.} 
HFU only reads each Gaussian coordinate and maximum scale in the coarse-grained filter, computing its projected position and coarse radius. 
These results are sent to the intersection test unit to determine if the Gaussian intersects with the current image tile. 
If the condition is met, HFU signals the codebook to decode the remaining parameters from the codebook buffer, as shown in \Fig{fig:arch}. 
The decoded parameters, along with the uncompressed parameters, are then fed into a FIFO for fine-grained filtering.

\paragraph{Fine-Grained Filter.} 
This step computes the exact radius computation and performs the intersection test.
Meanwhile, the fine-grained filter unit also computes the RGB values and conic matrix from decoded parameters, such as scales and rotation.
Based on the intersection result, HFU conditionally outputs a Gaussian RGB and conic results to the sorting buffer if this Gaussian passes the intersection test.

In our design, the coarse-grained filter largely reduces the computation (from 427 MACs to 55), compared to fine-grained filters.
In addition, the coarse-grained filter avoids decoding and reading unnecessary Gaussians to FIFO in HFU.

\vspace{-0.2cm}
\section{EVALUATION} 
\vspace{-0.2cm}



\subsection{Experimental Setup}
\label{sec:eval:exp}
\begin{table}[]
\centering
\caption{Configuration and Area}
\resizebox{\columnwidth}{!}{
\label{tab:area}
\begin{tabular}{lcc}
\toprule
Unit               & Configuration & Area [mm$^2$] \\ \midrule
Voxel Sorting Unit & 1 Unit & 0.06 \\
Hierarchical Filtering Unit & 4 Units    & 0.79       \\
Sorting Unit          & 2  Units    & 0.04       \\
Rendering Unit        & 64 Units    & 2.53       \\
SRAM (Input Buffer, Codebook, others)               & 355KB  & 1.95       \\ \hline
Total                &        & 5.37       \\ \bottomrule
\end{tabular}
}
\end{table}

\paragraph{Hardware Setup.}
Our accelerator consists of 1 VSU, 4 HFUs, 2 sorting units, and $4\times4\times4$ rendering units at 1GHz clock frequency.
Each HFU has 4 coarse-grained filter units (CFUs) and 1 fine-grained filter unit (FFU). 
The sorting unit and the rendering unit are identical to the bitonic sorting unit and the volume rendering unit in GSCore, respectively~\cite{lee2024gscore}.
The overall on-chip buffer consists of a double-buffered input buffer with a size of 16~KB, a 250~KB codebook buffer, and the remaining 89~KB to store intermediate data between stages.

Our architectural design is developed using an EDA process and synthesized with Synopsys and Cadence tools on TSMC’s 32~nm FinFET technology.
SRAM components are estimated by CACTI 7.0~\cite{cacti7} at 32~nm technology.
The DRAM model in our simulations is based on Micron’s 16 Gb LPDDR3 with 4 channels according to its datasheet~\cite{micronlpddr3}, with energy consumption data sourced from Micron System Power Calculators~\cite{microdrampower}.

\paragraph{Software Setup.} 
We configure a codebook with 4096 entries for scale, rotation, and DC, and a codebook with 512 entries for SH coefficients in \Sect{sec:algo:dc}. 
The voxel size is set to 2 for real world scenes and 0.4 for synthetic scenes. 
The hyper-parameter, $\beta$, in \Sect{sec:algo:fs} is set to 0.05.

\paragraph{Area.}
\proj design has a total of 5.37~mm$^2$ in \Tbl{tab:area}. 
The major area contributors are 4 HFUs (0.79~mm$^2$), 64 rendering units (2.53~mm$^2$), and on-chip buffers (1.95~mm$^2$).
Overall, our design has a similar area compared to GSCore (5.53~mm$^2$) which is scaled to 32~nm by DeepScaleTool~\cite{sarangi2021deepscaletool}.

\paragraph{Workloads.} We evaluate both synthetic and real-world datasets: Synthetic-NSVF~\cite{liu2020neural}, Synthetic-NeRF~\cite{mildenhall2021nerf}, Tanks\&Temples~\cite{knapitsch2017tanks} and Deep Blending~\cite{hedman2018deep}. 
Each scene is initially trained with a default $3\times10^4$ iterations, followed by $3\times10^3$ iterations of boundary-aware fine-tuning (\Sect{sec:algo:fs}) and $5\times10^3$ iterations of quantization-aware fine-tuning (\Sect{sec:algo:dc}).
We evaluated our method on three widely-adopted 3DGS algorithms, including original 3DGS~\cite{kerbl20233d}, Mini-Splatting~\cite{fang2024mini}, and LightGaussian~\cite{fan2023lightgaussian}.

\textbf{Baselines.}
We compare against two hardware baselines:  Nvidia GPU Orin NX~\cite{orinsoc} and a dedicated accelerator, GSCore~\cite{lee2024gscore}.
GPU performance and power consumption are obtained using built-in hardware measurements.
We implement GSCore based on its published specifications and validate our results against reported data.

\paragraph{Variants.} We evaluate three variants: 
\begin{itemize}
    \item \underline{w/o VQ+CGF} performs our fully-streaming algorithm without vector quantization and coarse-grained filtering.
    \item \underline{w/o CGF} performs our fully-streaming algorithm without coarse-grained filtering.
    \item \underline{\proj} is our full-fledged version including all optimization techniques.
\end{itemize}

\subsection{Rendering Quality}

\begin{table}[]

\caption{Quantitative comparison of rendering quality (PSNR $\uparrow$) across different datasets and algorithms. 
}
\label{tbl:PSNR}

\begin{tabular}{|p{2.2cm}@{}|@{ }c@{ }|@{ }c@{ }|@{ }c@{ }|@{ }c@{ }|@{ }c@{ }|@{ }c@{ }|}
\hline

\diagbox[width=2.4cm]{Algorithm}{Dataset} & Train  & Truck  & Playroom & Drjohnson & Lego   & Palace \\ \hline
3DGS       & 22.54 & 26.65 & 30.18 & 29.21 & 36.11 & 38.56 \\
Ours           & 22.52 & 26.61 & 30.27 & 29.07 & 36.02 & 38.52 \\
\hline
Mini-Splatting       & 21.49 & 25.19 & 30.32 & 29.23 & 36.20 & 39.00 \\
Ours           & 21.44 & 25.11 & 30.37 & 29.34 & 36.18 & 38.98 \\
\hline
LightGaussian       & 22.29 & 26.02 & 28.58 & 25.87 & 35.18 & 37.76 \\
Ours           & 22.32 & 25.89 & 28.47 & 25.79 & 35.15 & 37.68 \\
\hline
\end{tabular}

\end{table}

\Tbl{tbl:PSNR} shows a quantitative comparison of rendering quality between the original pipeline and our fully streaming pipeline across three 3DGS algorithms.
Overall, our fully streaming pipeline shows no accuracy degradation in visual quality.
On average, our pipeline drops the rendering quality by 0.04~dB.
Our results sometimes achieve higher PSNR than the original approach.
Our rendering results are also presented here: \href{https://streaminggs.github.io/}{link}.

\vspace{-3pt}

\subsection{Performance and Energy Efficiency}

\begin{figure}
    \centering
    \includegraphics[width=\linewidth]{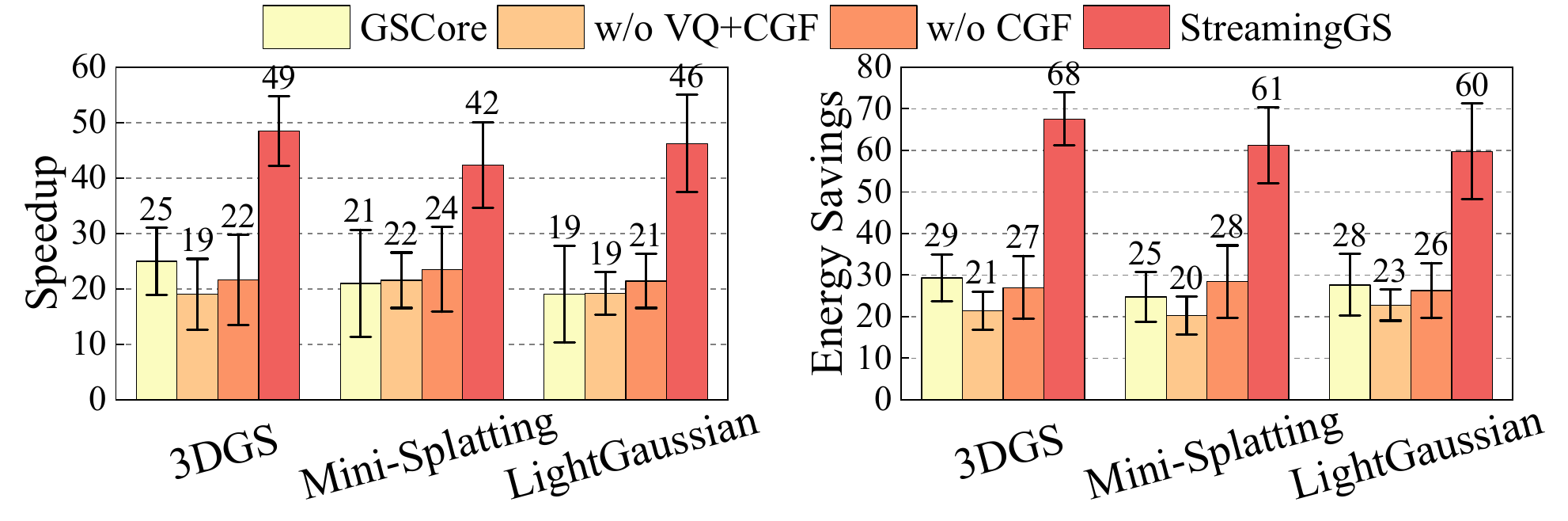}
    \caption{End-to-end speedup and energy savings of our variants over the GPU baseline. The numbers are averaged over four datasets and normalized to the GPU baseline.}
    \label{fig:speedup_energy}
\end{figure}

\Fig{fig:speedup_energy} illustrates the performance and energy savings of our design against three hardware baselines.
All values are normalized against the corresponding Orin NX. 
Both speedup and energy numbers are averaged across four datasets.

On average, \proj achieves 45.7$\times$ speedup against Orin NX. 
In comparison, GSCore can barely achieve 21.6$\times$ speedup.
Our result also shows that coarse-grained filtering affects the performance of \proj significantly.
Without coarse-grained filtering, the speedup drops from 45.7$\times$ to 22.2$\times$, while VQ has a minimal impact on the performance.
This is because VQ only affects the voxel streaming which is overlapped by the subsequent stages.

On energy saving, \proj achieves 62.9$\times$ energy reduction over Orin NX and 2.3$\times$ energy reduction over GSCore.
The primary energy savings are from DRAM traffic reduction and efficient early-stage irrelevant Gaussian filtering.
In addition, coarse-grained filtering and VQ contribute to 35.6$\times$ and 5.8$\times$ energy savings, respectively.


\vspace{-0.2cm}

\subsection{Sensitivity Study}
\label{exp:ss}

\begin{figure}[t]
\begin{minipage}[t]{0.48\linewidth}
        \centering
        \includegraphics[width=\textwidth]{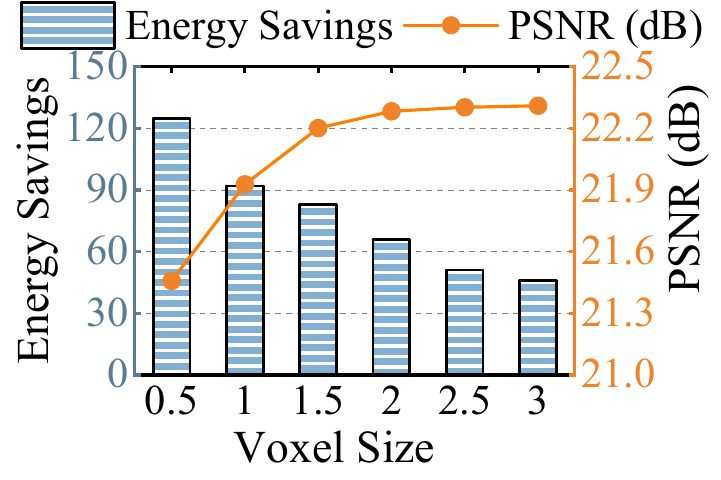}
        \caption{The sensitivity of energy efficiency and rendering quality on the original 3DGS to voxel size. We only evaluate on \textit{train} scene.}
        \label{fig:voxel_size}
    \end{minipage}
    \hspace{2pt}
    \begin{minipage}[t]{0.48\linewidth}
        \centering
        \includegraphics[width=\textwidth]{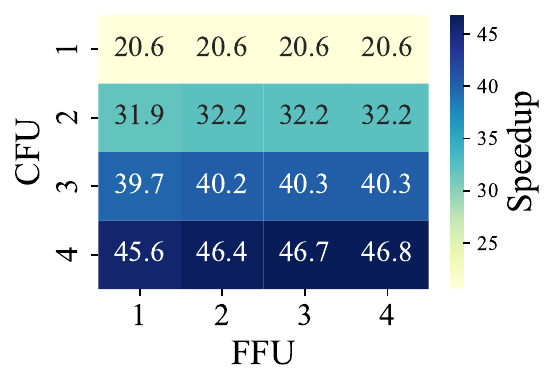}
        \caption{The sensitivity of speedup on the original 3DGS to the numbers of CFUs and FFUs. Only \textit{train} scene is evaluated.}

        \label{fig:heatmap}
    \end{minipage}
\end{figure}

\Fig{fig:voxel_size} shows the sensitivity of rendering quality and energy savings to voxel size on one real-world scene, \textit{train}.
All variants are retrained according to our training procedure in \Sect{sec:eval:exp}.
Generally, smaller voxel sizes result in more cross-boundary Gaussians, leading to incorrect depth ordering.
Our result shows that the rendering quality drops from 22.3~dB to 21.5~dB when the voxel size decreases from 2 to 0.5.
However, further increasing voxel size shows minimal impacts on rendering quality.
Meanwhile, larger voxels might potentially include more irrelevant Gaussians in every voxel, increasing the filtering workload and lowering energy efficiency. 
We empirically find a voxel size of 2 has a good balance between performance and rendering quality.

\Fig{fig:heatmap} illustrates the sensitivity of performance to the numbers of coarse-grained filter units (CFUs) and fine-grained filter units (FFUs) in HFU (\Sect{sec:arch:hfu}). 
We choose the same \textit{train} scene and normalize all performance numbers to the GPU baseline.
The results show that adding FFUs beyond the number of CFUs does not yield any speedup. 
In contrast, increasing the number of CFUs consistently boosts speedup, underscoring their importance in accelerating the filtering process. 
However, an excessive number of CFUs would lead to increased area overhead. 
Thus, we choose 4 CFUs with 1 FFU as our default configuration.


\vspace{-0.2cm}

\section{conclusion}

\vspace{-0.2cm}

This paper introduces \proj that rethinks the 3DGS pipeline and proposes a new computing paradigm to process Gaussian points.
Our fully streaming pipeline eliminates intermediate off-chip memory traffic and regularizes the off-chip accesses, all with a minimal but principle architecture augmentation.
Overall, our algorithm-architecture co-design demonstrates 45.7$\times$ and 2.1$\times$ speedup compared to mobile GPUs and the state-of-the-art 3DGS accelerator, respectively.

\section{ACKNOWLEDGMENT}
This work is sponsored by the National Natural Science Foundation of China (62472273, 62232015, 62402312).

\bibliographystyle{ieeetr}
\bibliography{ref}

\begin{thebibliography}{10}
\bibitem{visionprospec}
``{Apple Vision Pro screen refresh rate is up to 100Hz}.''
\end{thebibliography}

\end{document}